\documentclass{elsart}

\usepackage[dvips]{graphicx}

\usepackage{amssymb}

\begin{document}

\begin{frontmatter}



\title{G$^0$ Electronics and Data Acquisition\\(Forward-Angle Measurements)}


\author[IPN_Orsay]{D. Marchand,}
\author[IPN_Orsay]{J. Arvieux,}
\author[LPSC]{G. Batigne,}
\author[IPN_Orsay]{L. Bimbot,}
\author[CMU]{A. Biselli,}
\author[LPSC]{J. Bouvier,}
\author[UM]{H. Breuer,}
\author[CMU]{R. Clark,}
\author[IPN_Orsay]{J.-C. Cuzon,}
\author[IPN_Orsay]{M. Engrand,}
\author[LPSC]{R. Foglio,}
\author[LPSC]{C. Furget,}
\author[IPN_Orsay]{X. Grave,}
\author[LPSC]{B. Guillon,}
\author[IPN_Orsay]{H. Guler,}
\author[UM,UIUC]{P.M. King,}
\author[LPSC]{S. Kox,}
\author[CMU]{J. Kuhn,}
\author[IPN_Orsay]{Y. Ky,}
\author[CMU]{J. Lachniet,}
\author[IPN_Orsay]{J. Lenoble,}
\author[LPSC]{E. Liatard,}
\author[UM]{J. Liu,}
\author[LPSC]{E. Munoz,}
\author[LPSC]{J. Pouxe,}
\author[LPSC]{G. Qu\'em\'ener,}
\author[CMU]{B. Quinn, }
\author[LPSC]{J.-S. R\'eal,}
\author[LPSC]{O. Rossetto,}
\author[IPN_Orsay]{R. Sellem,}

\address[CMU]{Department of Physics, Carnegie Mellon University, Pittsburgh, PA 15213, USA}
\address[IPN_Orsay]{Institut de Physique Nucl\'eaire d'Orsay (UMR8608), CNRS/IN2P3, Universit\'e Paris Sud, Orsay, France}
\address[LPSC]{LPSC, Universit\'e Joseph Fourier Grenoble 1, CNRS/IN2P3, Institut National Polytechnique de Grenoble, Grenoble, France}
\address[UIUC]{Loomis Laboratory of Physics, University of Illinois, Urbana IL 61801, USA}
\address[UM]{Department of Physics, University of Maryland, College Park, MD 20472, USA}

\begin{abstract}
The G$^0$ parity-violation experiment at Jefferson Lab (Newport News, VA) 
 is designed to determine the contribution of strange/anti-strange quark pairs to the intrinsic properties 
 of the proton. In the forward-angle part of the experiment, the
 asymmetry in the cross section was measured for $\vec{e}p$ elastic 
scattering by counting the recoil protons corresponding to the two beam-helicity states. 
Due to the high accuracy required on the asymmetry, the G$^0$
 experiment was based on a custom experimental 
setup with its own associated electronics and data acquisition (DAQ)
 system. Highly specialized time-encoding electronics provided time-of-flight 
spectra for each detector for each helicity state. More conventional 
electronics was used for monitoring (mainly FastBus). The time-encoding electronics 
 and the DAQ system have been designed to handle events at a mean rate
 of 2 MHz per detector with low deadtime and to minimize
 helicity-correlated systematic errors. In this paper, we outline the general architecture and the 
 main features of the electronics and the DAQ system dedicated to G$^0$ forward-angle measurements.
\end{abstract}

\end{frontmatter}

\section{Introduction}
\label{Intro}
The G$^0$ parity-violation experiment described in references \cite{G0_PAVI04} and \cite{G0NIM} was performed in hall C at 
the Jefferson Laboratory (Newport News, VA). The goal of this experiment was to measure the strangeness content of the 
nucleon, determining the strange electric and magnetic form factors of the proton. Combining forward- and backward-angle 
measurements (Rosenbluth separation), the G$^0$ experiment will be able to separate the strange electric form factor from the 
strange magnetic one for specific Q$^2$ values (at least Q$^2$=0.23\ (GeV/$c$)$^2$ and Q$^2$=0.64\ (GeV/$c$)$^2$). The 
first data-taking performed in 2004 was dedicated to forward-angle measurements which determined a linear combination of the 
strange electric and magnetic form factors for each of eighteen Q$^2$-bins, extending from 0.1 to 1.0 (GeV/$c$)$^2$. The 
results, published in reference \cite{G0_FW_Results}, indicate nonzero, Q$^2$-dependent, strange-quark contributions to the 
charge and/or current densities of the proton. In the forward-angle mode, measurements were made of the counting rate 
asymmetry for recoil protons from the elastic scattering of longitudinally polarized electrons from unpolarized protons. This 
asymmetry, arising from the parity violation property of the weak interaction (Z$^0$ boson exchange), is of the order of 10$^{-6}$ 
to 10$^{-5}$ within the Q$^2$ range of interest. Measuring such small asymmetries required counting statistics of the order of 
10$^{13}$  events which led to specific requirements such as a high intensity electron beam, a thick, high-power LH$_2$ target, 
a large acceptance detector system including a spectrometer and a dedicated electronics system allowing time-of-flight 
measurements to isolate elastic scattering from inelastic processes.  The requisite data rate excluded the possibility of 
event-by-event readout, necessitating spectrometer optics which separate the elastic events without the use of tracking 
detectors. As the angles of elastically scattered electrons are very close to the incident beam (7$^{\circ}$ to 21$^{\circ}$), it was 
preferable to detect the recoil protons between 52$^{\circ}$ and 76$^{\circ}$.

The G$^0$ experimental setup  was designed to be used for both forward- and backward-angle measurements. The 
backward-angle measurements, performed in 2006-2007, required the rotation of the spectrometer by 180$^{\circ}$ and the 
addition of various detectors (scintillators and Cerenkov counters) in order to detect scattered electrons from large angle 
$\vec{e}p$ elastic scattering and quasi-elastic electron-deuteron scattering. A description of the experimental set-up for 
backward-angle measurements including its specific electronics can be found in reference \cite{G0NIM}.

For the forward-angle measurements, a brief description of the experimental apparatus and the requirements of the G$^0$ beam 
structure are given in the next section. A description of the data acquisition and the trigger systems is presented in section 
\ref{section:DAQ}. The electronics, consisting of a monitoring system (FastBus) and time-encoding electronics, is  outlined in 
section \ref{section:electronics} while the details of the two time-encoding sub-systems are given in sections \ref{section:NA} and 
\ref{section:FR}.
  
\section{G$^0$ Experiment for Forward-Angle Measurements}
\label{section:overview_exp}
\subsection{Overview}
\label{sub:overview}
\begin{figure}[h]
\begin{center}
\includegraphics[width=12cm]{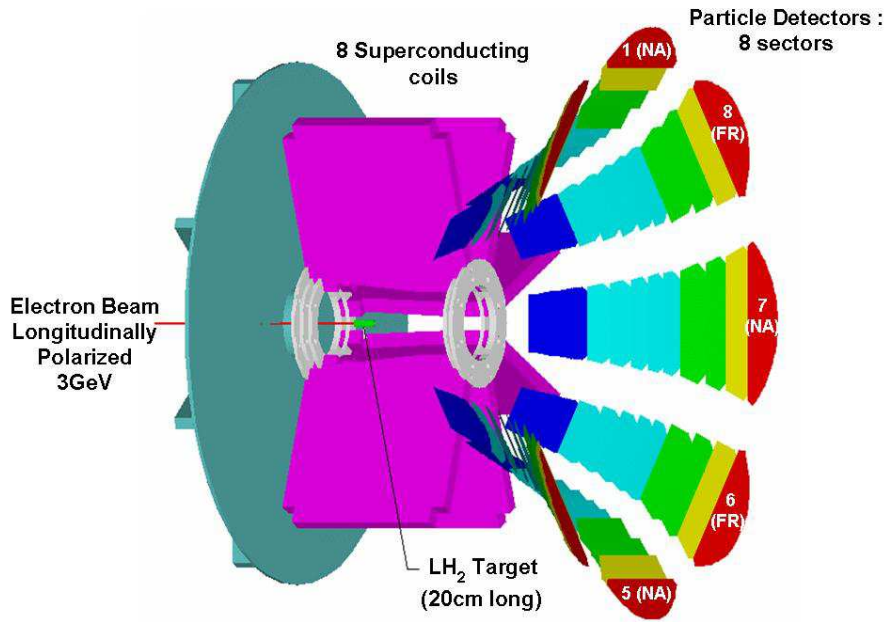}  
\caption{The G$^0$ experimental setup dedicated to forward-angle measurements. It consists of a 20 cm liquid hydrogen target 
and a toroidal Superconducting Magnet System (SMS) followed by 128 pairs (front and back) of plastic scintillators grouped into 
eight sectors (octants). Odd numbered octants have been built by the North American (NA) part of the collaboration and the even 
numbered octants by the French (FR) part. Two of the superconducting coils and one octant have been removed for clarity.}
\label{fig:G0setup}
\end{center}
\end{figure}

The forward-angle part of the G$^0$ experiment measured asymmetries in $\vec{e}p$ elastic scattering by detecting the recoil 
protons. As illustrated in figure \ref{fig:G0setup}, a longitudinally polarized 3 GeV electron beam struck a liquid hydrogen target 
located on the axis of a toroidal superconducting magnet. Recoil elastic/inelastic protons and other positively charged particles 
are deflected by the magnetic field to focal plane detectors grouped into eight sectors (referred to as octants). Each sector is 
instrumented by 16 pairs (two layers) of plastic scintillators.  
The identification of recoil elastic protons relies on time-of-flight measurements relative to the beam pulse arrival time in the 
target. This required a custom pulsed beam structure to provide a start time, rather than the nominal continuous beam.

\subsection{Electron Beam}
\label{sub:electron_beam}
To allow time-of-flight measurements, for forward angle G$^0$ running, the beam had  a specialized time structure of 31.1875 
MHz, 1/16 of the standard beam pulse frequency of 499 MHz provided to each of the three halls. Therefore electron beam pulses 
were delivered with a period of approximately 32 ns to Hall C at an average beam current of 40 $\mu$A. With this beam 
structure, each pulse must carry 16 times more electrons than for the same average current in 499 MHz operation.

Effects due to the 60 Hz line frequency were averaged out by choosing a data-taking period of $\frac{1}{30}$ second (referred to 
as one MacroPulSe, or MPS). No data were taken during an additional  period of 500 $\mu$s  following each MPS. At the 
beginning of this period the helicity of the beam was reversed, if required, and the remainder of the period allowed time for 
signals to stabilize.

The macropulses were grouped into quartets. Within each quartet the helicities of the four consecutive macropulses followed 
one of the two allowed patterns, either $+--+$ or $-++-$. The choice of which pattern to use for a particular quartet was made 
pseudo-randomly to prevent any long-term periodicity of the helicity sequence. These patterns reduce sensitivity to any linear 
drifts within a quartet and also ensure that asymmetry determinations can be made quartet-by-quartet.  The MPS signal was sent 
to all sub-systems through the ``trigger supervisor'' described in section \ref{section:DAQ}. The helicity sequence for each quartet 
was buffered at the source and reported to the experiment only after a delay of two quartets.  This precaution prevented the 
helicity-related signals from having any influence on the electronics which might otherwise induce a false asymmetry.

A custom-built system \cite{Musson_box} provided a very stable beam-arrival signal, phase-locked to a 1497 MHz resonance 
cavity installed just upstream of the G$^0$ target. This signal (referred to as Y$_0$) served as the start signal for the 
time-of-flight measurements and was sent to both time-encoding electronics sub-systems and to the monitoring system 
(FastBus) as described in section \ref{section:electronics}. Time-of-flight measurements were made relative to Y$_0$, thus 
avoiding the possibility of introducing a helicity-dependent bias to the time-of-flight measurements due to helicity-dependent 
transit time of the beam through the accelerator.  A systematic shift of the order of ten femto-seconds could introduce an 
unacceptable false asymmetry. Time-of-flight measurements with respect to beam-arrival time eliminated that dependence on 
transit time.

\label{section:Y0_introduced}
While the 1497 MHz R.F. signal from the resonance cavity provides an accurate reference to phase-lock the Y$_0$ signal, the 
48-fold ambiguity in the time of the Y$_0$ was resolved using a 31.1875 MHz oscillator phase-locked to the signal from a 
stripline beam pickup located near the cavity. The latter signal was used to reproducibly select a unique phase of Y$_0$ with 
respect to the 1497 MHz cavity signal. The response time for the signals to accommodate to a change in beam timing was 100 
$\mu$s.  Thus, a possible transit time shift due to beam-helicity reversal would have been accommodated during the 
helicity-stabilization period before resumption of data-taking.

\subsection{Apparatus}
The 20 cm long $G^0$ liquid hydrogen target was designed to minimize density fluctuations with up to 250 Watts of beam power 
deposited in the target. The target could be removed from the beam for diagnosis purposes. Details on the design and 
performances of the target can be found in references \cite{G0NIM}, \cite{G0target} and \cite{these_Silviu}. This target was also 
used for backward-angle measurements with liquid Hydrogen and Deuterium.

Detection of scattered particles was based on a superconducting magnetic spectrometer \cite{SMS_PAVI04,G0NIM} (referred to 
as SMS, figure \ref{fig:G0setup}). It focused elastic recoil protons through collimators onto a Q$^2$-segmented Focal Plane 
Detector array (plastic scintillators). The toroidal superconducting magnet of the spectrometer is composed of eight 
superconducting coils surrounding the beam axis. These coils split the detector system into octants. The SMS cryostat has a 
mean diameter of about 4 m and an axial length of 2 m. The magnetic  field integral is $\int Bdl \approx $1.6 T$\cdot$m. The 
angular deflection in the magnetic  field ranges from 35$^{\circ}$ to 87$^{\circ}$. The azimuthal acceptance of each octant, about 
22$^{\circ}$, was accurately defined by horizontal collimators. Vertical collimators limited the polar-angle acceptance to that 
corresponding to elastic scattering over the $Q^2$ range of interest, reducing counting rates from background. The resulting total 
solid angle was 0.9 sr. 

The SMS was designed such that elastic recoil protons having the same momentum are focused at the same point in the focal 
plane independently of the interaction point in the target along the beam axis.

\label{section:Q_bins}
Each octant was instrumented with 16 Focal Plane Detectors (FPDs), each consisting of two layers made of plastic scintillator 
paddles (Bicron BC408) shaped and segmented to detect elastic protons corresponding to  narrow bands of Q$^2$. The FPDs 
in each octant were numbered sequentially, with the first FPD the closest to the electron beam line and the sixteenth the farthest 
one out. This is also the direction of increase for the Q$^2$ bins until a turn-around point located at detector 15. The detector 
system was designed such that each of the first 13 FPDs corresponds to a specific Q$^2$ bin. The next two FPDs (14 and 15), 
due to the optics of the spectrometer covered a larger Q$^2$ range, because of the kinematic turn-around point. The last FPD 
(detector 16) by design did not observe any elastic events and was used for background studies.

The construction of the detectors was shared between the North American (NA) and the French (FR) parts of the collaboration 
so that each part built four octants with its associated time-encoding electronics. As shown in figure \ref{fig:G0setup}, the NA and 
FR octants were mounted alternately around the beam axis to maximize azimuthal symmetry. The two-layer scintillator paddle 
pair of an FPD, referred to as the front and the back scintillators, were separated by a 3 mm thick sheet of polycarbonate (NA) or 
Aluminum (FR). These sheets reduced the rate of background coincidences caused by interactions in one scintillator producing 
secondaries which enter the other. Each end of each scintillator paddle was connected to a Photomultiplier Tube (PMT) through 
a light guide. The NA FPDs used Photonis XP2262B 12-stage PMTs with custom-built Zener-diode bases while the FR FPDs 
use Photonis XP2282B 8-stage PMTs with actively stabilized custom-made bases. To increase the PMT life-time, the PMTs 
were run at low gain and the signals were amplified.  The FR bases include a pre-amplifier providing an amplification factor of 
20. The amplification factor for the NA tubes was 25, provided by modified commercial PMT amplifiers.

To reduce background, particle detection was based on front-back coincidences in the two scintillator layers of each FPD. 
Particle identification (separation of recoil elastic protons from inelastic protons and pions)  relied upon Time-of-Flight (ToF) 
measurements. The ToF spectra were accumulated by highly specialized time-encoding electronics described in section 
\ref{section:electronics}. For cross-checking purposes and to control systematic errors, the G$^0$ experiment benefits from two 
different designs of time-encoding electronics: one associated with the four NA octants and the other associated to the four FR 
octants. Thus, each set of data provided an independent complete measurement. The two results could be checked for 
consistency before being combined into the final result for the entire set of octants.  In addition to the custom time-encoding 
electronics, conventional FastBus electronics (ADCs and TDCs) were used to acquire and store complete event-by-event data 
for a small fraction of beam pulses (less than 0.01\%). 

\section{Data Acquisition System}
\label{section:DAQ}
\begin{figure}[h]
\begin{center}
\includegraphics[width=10cm]{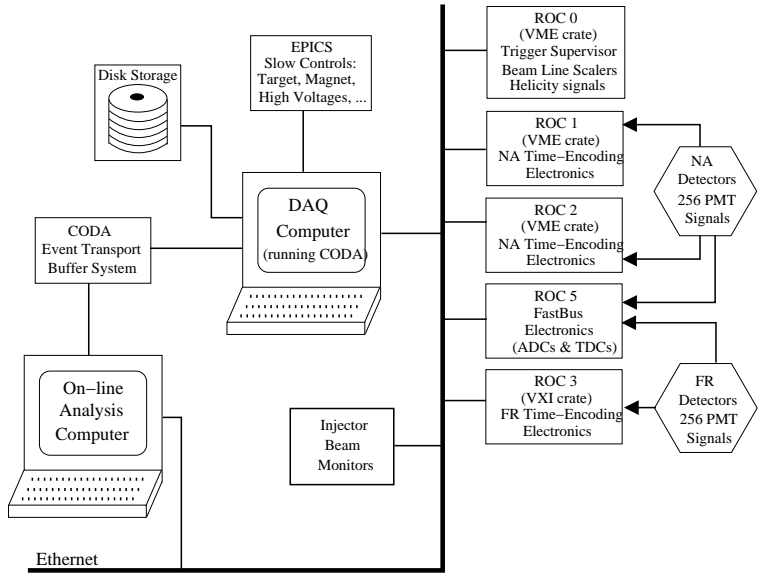}  
\caption{Schematic of the G$^0$ Data Acquisition system. ROC stands for Read-Out Controller.}
\label{fig:G0DAQ}
\end{center}
\end{figure}

The G$^0$ data acquisition was driven by the CODA (CEBAF On-line Data Acquisition) \cite{coda} system developed at JLab. 
The triggering and the event control were performed by the Trigger Supervisor module linked to each subsystem/crate using 
Trigger Interface modules. The Trigger Supervisor served as the interface between the experiment-specific triggering and the 
data acquisition system. The different trigger sources were input to the Trigger Supervisor. A schematic of the G$^0$ Data 
Acquisition system is presented in figure \ref{fig:G0DAQ}.

During production runs, the data readout of both time-encoding sub-systems was performed at the MPS rate (roughly 30 Hz). At 
the end of each MPS, during the ``helicity-stabilization'' period, the data were stored in memory buffers to allow them to be read 
out during the next MPS. After each MPS, the beam parameters and the data from each time-encoding sub-system were first 
transferred to a Read Out Controller (ROC) and then gathered with the other ROCs as a single event by the ``Event Builder" in 
the DAQ computer. For the specialized time-encoding electronics, time-of-flight spectra corresponding to each MPS were stored. 
This allows an off-line cut on the data quality for each MPS before calculating asymmetries quartet by quartet. 

In order to detect any possible 60 Hz noise, an ``over-sampling'' running mode at 120 Hz was performed occasionally. In this 
running mode,  the NA Time-Encoding Electronics (TEE) were stored and read out four times during each MPS, once after each 
1/120 of a second.  The French TEE were still read out at 30 Hz, after each MPS. For the French TEE, the subdivision of the 
data into subsets collected in four sub-periods of the MPS was made by means of the front-end Digital Signal Processors of 
each mother board. During 120 Hz acquisition, only TEE and beam parameter scalers were read out. A Fourier analysis of the 
120 Hz data (ToF spectra, positions and angles of the beam, beam charge) performed after each 120 Hz run determined that the 
specifications regarding 60 Hz noise have been achieved.

The readout of the FastBus data (monitor events) was triggered by a prescaled copy of the Y$_0$ signal. This was first 
prescaled in hardware to provide an input rate of about 500 kHz. For typical running conditions, the rate was then reduced to 
100-200 Hz by applying a software prescale factor within the trigger supervisor. More details on the G$^0$ trigger system can be 
found in reference \cite{G0NIM}.

Another set of slow control data was also recorded to control and monitor the experimental apparatus. These events contain 
information such as detector high voltages, beam characteristics (charge, positions, ...), target and SMS parameters.

\section{Electronics}
\label{section:electronics}
\begin{figure}[h]
\begin{center}
\includegraphics[width=15cm]{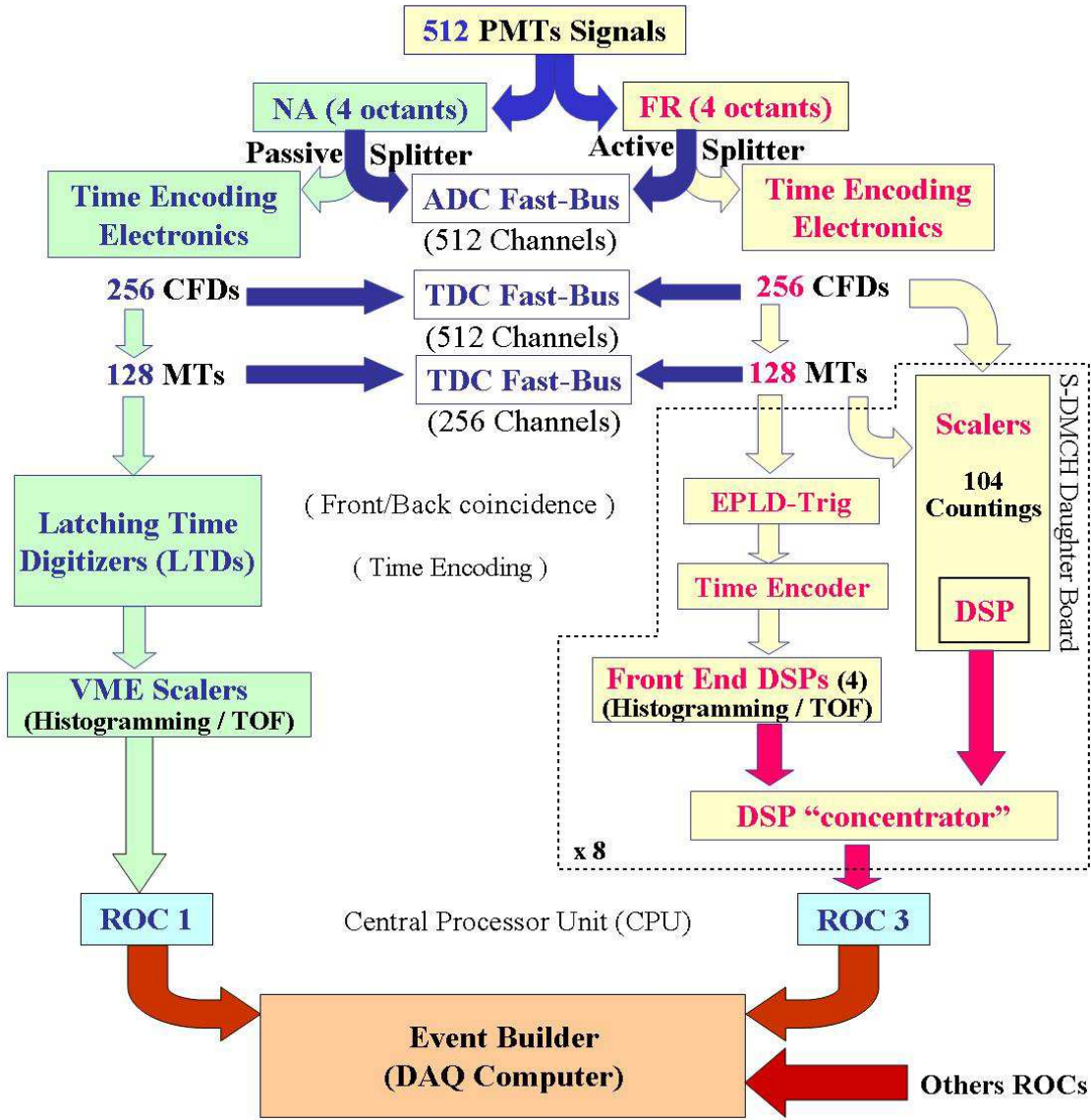}  
\caption{General scheme of the G$^0$ electronics chain from PMT signals to full-event buffers.}
\label{fig:architecture}
\end{center}
\end{figure}

At a mean rate of 2 MHz per detector, it is impossible to store full event-by-event information (timing and amplitude of the PMT 
signals). Instead, it was chosen to accumulate, in hardware, histograms of the time-of-flight information provided by the 
time-encoding electronics. For each FPD the mean arrival time is stored for complete events, which have a front-back 
coincidence. No information is stored using the TEE concerning, for example, time differences between signals from the two 
PMTs of one scintillator. In order to monitor such detector information, conventional FastBus electronics is used to store full 
event information for a few beam pulses out of every $10^5$. Figure \ref{fig:architecture} summarizes the whole electronics 
chain from PMT signals to full-event buffers. 

\subsection{Monitoring system}
\label{subsection:MSyst}
The monitor electronics consists of FastBus ADCs (LeCroy 1885F) and TDCs (LeCroy 1875A). After passing through about 36 
m of RG58 cable in the experimental hall, the PMT signals were carried up to the electronics/DAQ area on 107 m of RG8 cable. 
In the electronics area, the design allows all signals to go to the TEE but also presents them to the FastBus monitor electronics 
so a small subset can be sampled on an event-by-event basis. Splitter modules route part of the PMT signal to FastBus ADCs 
while the other part of the signal goes to the Constant Fraction Discriminators (CFDs) which are the front-end of the TEE. The 
French PMT signals enter active splitters (gain 1) so that the signal sent to the TEE remains unchanged and the signal sent to 
FastBus ADCs is a copy of the original signal. The North American PMT signals enter passive impedance-matching 2:1 splitters 
which send the smaller fraction of the signal to FastBus ADCs. The larger fraction of the split signal goes to the input of 
commercial CFD modules. In order to achieve good time resolution, the two CFD signals from each scintillator are meantimed in 
hardware in each TEE sub-system. Copies of the CFD and MeanTimer (MT) outputs are also routed to FastBus TDCs. For an 
arbitrarily selected set of beam pulses, the ADCs were gated and the TDCs were started and all resulting FastBus information 
was read out. The combination of ADC and TDC information provides full event-by-event information for this small subset of 
events. These FastBus events allow the study of correlations (in pulse size and/or in time) for individual detectors and for 
relationships between detectors.  The stored FastBus events are used to study the behavior of the meantimers, to measure 
timing resolution, to monitor singles rates, to measure attenuation in the scintillators, and to perform other checks of the 
apparatus.  Some of the measured quantities, such as rates of single-, double-, or triple-PMT coincidences are also used to 
quantify deadtime corrections to apply to the TEE data. For some low-luminosity calibration runs a fast-clear signal to all ADCs 
and TDCs was used to discard those FastBus events for which the CFD pattern indicated the lack of any charged particles of 
interest in the detectors.
 
\subsection{Time-Encoding Electronics}
\label{subsection:TEE}
The challenge in the design of the TEE is to accept all the four-fold coincidence events 
(complete events, requiring both PMTs for both front and back scintillators) at a mean rate of 2 MHz per detector and to 
accomplish the time-of-flight measurement and the subsequent histogramming. For cross-checking purposes, the G$^0$ 
experiment benefits from two different designs  for TEE. The NA electronics, designed by Carnegie Mellon University 
(Pittsburgh, PA), is mostly based on modular electronics units whereas the French electronics, designed by IPN Orsay, is fully 
integrated. The advantage of such divergent paths was that there was little likelihood that both techniques would be susceptible 
to the same systematic errors.

Both time-encoding sub-systems are based on similar objectives in processing the front-end signals. After passing through the 
splitters, the PMT signals are first sent to Constant Fraction Discriminators (CFDs) to reject low energy background. Since offline 
walk corrections are prohibited by the lack of event-by-event data, CFDs have been chosen to provide
good time resolution over a large dynamic range as the firing time is, in principle, independent of the amplitude of the input 
signal. Then, to achieve good time resolution, the timing signals from opposite ends of each
scintillator are meantimed in hardware. This ensures that  the encoded
time is almost independent of the hit location on the scintillator paddles. The
MT signals are provided to a logic device which makes the coincidence
between the front and the back scintillator signals. The timing of the
coincidence events is encoded by time digitizers relative to the
Y$_0$ signal, to determine the time-of-flight. Finally, according to their timing, each coincidence
event increments the corresponding time bin in a
time-of-flight spectrum built by means of VME scalers (NA) or digital
signal processors (FR).

The two electronics designs include common features to control and monitor the dead time. In order to make dead time more 
deterministic, Next Pulse Neutralization (NPN) is enforced, meaning that encoding is disabled during the next beam pulse after 
a coincidence event occurs.  Another feature which has been implemented in both sub-systems is the 
so-called ``buddy'' system which is used mostly as an alert to monitor beam charge 
fluctuation and detect any helicity dependence of dead time. In this context, similar detectors 
(symmetric to each other at 180$^{\circ}$ w.r.t. the beam axis) have been paired as ``buddies'' and separate count is kept of hits 
occurring on a detector while the electronics of its buddy detector are busy. Missed events due to dead 
time may be expected to scale like these ``buddy'' events.
Besides the ``buddy'' system, the French electronics provides time-of-flight spectra of ``buddy'' 
events (referred to as ``differential buddy'') which takes into account the NPN. More details on the 
differential buddy method and its outcomes can be found in reference \cite{Diff_Buddy}.

The characteristics of each of the two time-encoding sub-systems are given in the next two sections. 

\section{North-American Time-Encoding Electronics}
\label{section:NA}
\subsection{Overview}

The electronics which was developed to read out the `North-American' octants was based on 
a highly modular approach using the simplest  design consistent with the requirement of 
encoding time-of-flight spectra at rates averaging about 2 MHz per detector.

Discrimination and meantiming are performed in separate modules. The resulting meantimed signals for each front scintillator 
and the corresponding back scintillator are then presented to the time-encoding
board, called a Latching Time Digitizer (LTD).  The LTD demands a front-back 
coincidence and, if it is present, encodes the time-of-flight of the front 
meantimer as a bit pattern, as described in section \ref{subsubsection:TEmodule}. The bit patterns for all beam pulses within an 
MPS are then accumulated in channels of scaler modules at instantaneous  rates up to 15.6 MHz.  The main role of the LTD is to 
encode time-of-flight as scaler channels.  At the end of an MPS, the accumulated scalers contain
the information needed to construct the time spectrum.  As described below, the
time spectra accumulated during one MPS can be rapidly captured, allowing accumulation of the time spectra for the next MPS 
to begin while the spectra from the earlier MPS are being read out by the DAQ.

In addition to the series of modules through which pulses pass to be encoded 
as time spectra, a set of auxiliary modules are needed to generate clock 
trains to the LTDs.  These auxiliary modules provide synchronization and time
reference to the LTDs.

\subsection{Signal Processing}
\label{sub:Signal_Processing}

When the PMT signals enter the electronics area the passive 2:1 splitter, discussed in section \ref{subsection:MSyst}, sends the 
larger output signal to a LeCroy 3420 Constant Fraction Discriminator. At the high rates characteristic of this experiment, it was 
found that care was needed to avoid unacceptable loss of data due to the effect of sub-threshold signals distorting the pulse 
shape for above-threshold events.  This effect was resolved by careful choice of the internal delay and fraction parameters of the 
CFD.  As indicated in figure \ref{fig:architecture}, the differential ECL output signal of the CFD was passively split to go to a 
meantimer and to a FastBus TDC  (LeCroy 1875A) to be sampled event-by-event for monitoring purposes.

\begin{figure}[h]
\begin{center}
\includegraphics[width=12cm]{{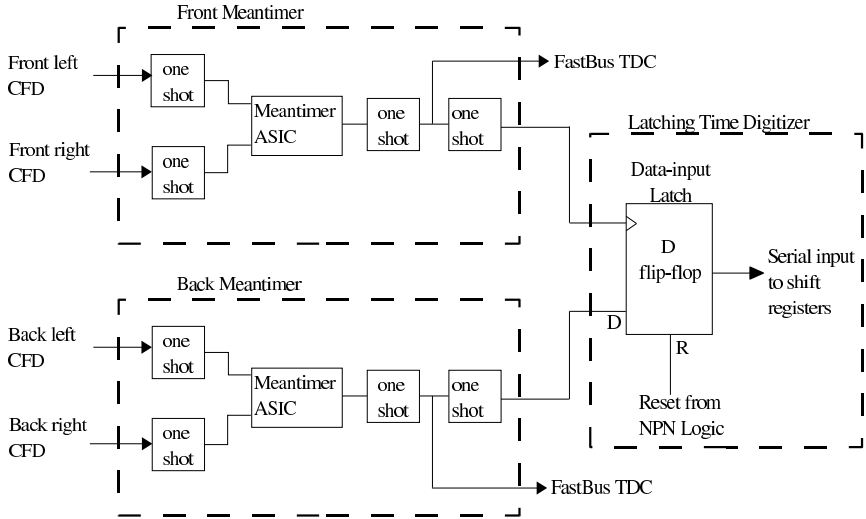}}   
\caption{A schematic representation of the path for NA detector
  signals (from a single FPD) through the meantimer modules and into
  the data-input latch of the LTD.  The wedge on the D flip-flop
  indicates the Clock input while letters mark the inputs for Data and Reset.}
\vspace{0.7cm} 
\label{fig:NA_fig}
\end{center}
\end{figure}

Hardware meantiming was used to render the time of the signals relatively insensitive to the position along the length of the 
scintillator at which the proton crossed. Custom meantimers were constructed at Carnegie Mellon University based 
on a meantiming ASIC (Application-Specific Integrated Circuit) developed at LPSC-Grenoble \cite{MT_LPSC}.  This ASIC is 
based on two shift-register pipelines which clock the incoming PMT signals in opposite directions. The meantime output fires 
when two signals overlap at the same position in the two counter-propagating pipelines. This simple principle of operation 
avoids recovery effects, such as long deadtime or time distortion, which are sometimes
encountered in other meantimer designs.  The operations performed by the meantimer board are represented schematically in 
figure \ref{fig:NA_fig}. Features added by the meantimer 
board include reduction of accidental overlap of unrelated pulses by replacing 
the input pulses from the CFD with short one-shot pulses of about 2 ns fed into the meantimer ASIC. 
The meantimed output pulses also go through a pair of consecutive one-shots to 
allow separate selection of meantimer output pulse width and deadtime, subject 
only to the constraint that the deadtime exceed this output width.  The outputs 
from the first of these one-shots were sent to FastBus TDCs (LeCroy 1875A) 
for event-by-event monitoring.   The narrower pulse from the second one-shot was sent to the LTDs to form the front-back 
coincidence. The meantimer boards also provided temperature monitoring and compensation for shifts in clocking speed of the 
ASIC with temperature.

\subsubsection{Clock Gating and Distribution}
\begin{figure}[h]
\begin{center}
\includegraphics[width=11cm]{{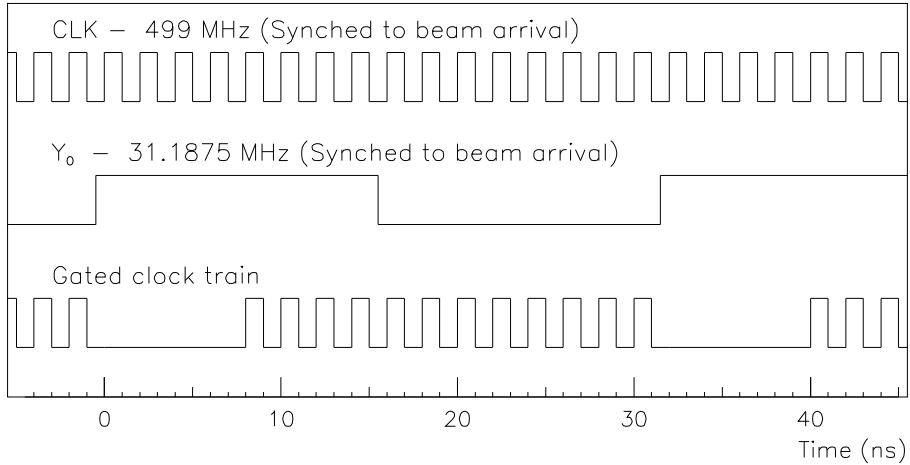}}   
\caption{Timing diagram showing clock trains generated based on the CLK and Y$_0$ signals 
described in the text.}
\vspace{0.7cm} 
\label{fig:clock_train}
\end{center}
\end{figure}

The NA time-encoding modules, described in the next section, use 
externally-generated clock trains as the basis for their time measurements.
Generation of clock trains for accurate time-of-flight measurements is based on 
two signals synchronized to beam arrival at the target.  These signals are 
labeled as CLK and Y$_0$ in figure \ref{fig:clock_train}.  CLK has a frequency
of 499  MHz, while Y$_0$ (as described in section \ref{section:Y0_introduced}) 
has a frequency of 31.1875 MHz, the frequency of the 
beam pulses.  The CLK signal (like the Y$_0$) is phase-locked to the
arrival time of the pulses at the target rather than being aligned relative to 
the accelerator's R.F. or to the injection. 
CLK is generated from the JLab 499 MHz master oscillator signal by phase-shifting it
to align with Y$_0$. 

Figure \ref{fig:clock_train} illustrates how the CLK and Y$_0$ are used in a 
custom clock-gating board to produce trains of 12 clock pulses (with the 
period of the 499 MHz CLK pulses).  The clock-gating board simply gates off 
four CLK pulses with a gate synchronized to Y$_0$.  The width of the gate and
its offset relative to Y$_0$ are adjustable by on-board switches to compensate for cable delay variations.

A synchronization pulse, SYNC, is also generated
by the clock-gating board between clock trains.  The clock train and SYNC are 
distributed by twinax cable to custom 9-fold signal duplication boards and then 
fanned out, on the same type of cable, to the LTD boards. Both signals are distributed as complementary ECL to improve noise 
immunity.

\subsubsection{Time Encoding Module}
\label{subsubsection:TEmodule}

The time-encoding boards (LTDs) enforce the 
requirement of a front-back coincidence between the scintillators. As indicated schematically in figure \ref{fig:NA_fig}, a D 
flip-flop is clocked by the front scintillator meantime while the meantimer pulse
for the back scintillator is presented to the D input.  Thus the time at which 
this data-input latch is set is determined by the front scintillator while the tolerance 
of the coincidence is set by the width of the pulse for the back
meantimer, which was set to 10 ns.

The principle of operation of the LTDs provides a very simple method of
accumulating time spectra for data rates of several MHz.  In fact the LTD boards
were tested successfully with rates of several tens of MHz.  The time-encoding
behaved as expected even at those rates and deadtime corrections were
correctly predicted although deadtime corrections began to become large, as
expected.  No measurable shift of the edges of the time bins was observed for pulsed data
rates ranging from a few kHz to 31.1875 MHz. An upper limit on such variation was set at a few ps per MHz.

As described below, the LTDs work by encoding each time spectrum as a simple pattern of 24 bits which can then be
presented to scaler inputs.  The scalers accumulate the bit patterns and thus
store information from which a time spectrum can be easily unfolded.  Only this spectrum is stored, so all correlations within one 
event are lost. The bit pattern is generated by presenting the output of the data-input latch to the serial input of a 
previously-cleared shift register.  The shift register is clocked
by the gated clock train described above and shown in figure \ref{fig:clock_train}. If the data-input latch is set during any clock 
train, a series of ``ones'' will be clocked into the shift
register.  Earlier hits within a clock train will result in a longer train of
``ones'' being clocked in.  Since the data-input latch is not reset until after
the end of the clock train, the data clocked into the shift register will have
a very simple format with a single zero-to-one transition encoding the time of
the hit.  Every 32 ns, after each clock train, the parallel outputs of
the shift register are strobed to latches which present them to scaler inputs.
 The scalers are separate VME modules whose differential inputs are driven via
twisted pair ribbon cables driven by these latches.  During the period between clock trains the shift registers are also reset. The 
strobe of the latch and the reset of the shift registers are timed relative to the last pulse of the clock trains. 
The SYNC signal, which arrives in the period between clock trains, is used to detect and reset any loss of synchronization which 
might occur if a clock train were incorrectly gated or if the electronics made an error in counting the pulses in the clock train. The 
reset of the data-input latch is described below. 

Finer time resolution is achieved by presenting each data-input latch to two
parallel shift registers, one of which is clocked by the leading edges of the
gated clock train while the other is clocked by the trailing edges.  One or
the other of the two shift registers is thus clocked each nano-second during the clock
train.  Treating the two parallel shift registers as a single inter-leaved
sequence of 24 bits, the position of the zero-to-one transition then encodes the
time of the data with a time bin of 1 ns.   
Differences between sequential bits will then be non-zero only for the time bin in which the hit occurs. 
Similarly, after these bit patterns have been accumulated in scalers,
differences between sequential scaler channels (arranged in the order of sequential bits) represent the total number of
hits which occurred in the corresponding 1 ns time bin in all clock trains for
which data were accumulated.  Time spectra can thus be determined, for each MPS, with the scaler values being read out only 
after each MPS. The scaler-channel differences span a 23 ns range of the 32 ns between beam pulses. This yields a TOF 
spectrum covering a 23 ns period of interest, with 1 ns time resolution. The 24th bin keeps count of events which occur in the 
several nano-seconds between the reset of the data-input latch and the beginning of the next clock train.  

\subsubsection{Scalers and DAQ Interface}

The scalers that capture the time spectra are custom-built 32-channel VME scaler boards which
were designed by LPSC-Grenoble \cite{VME-scalers} based on a scaler
ASIC developed there. Functionally they are used as latching scalers in this application.  Data are 
accumulated for the duration of an MPS ($\frac{1}{30}$ of a second) during which the beam helicity is fixed. 
Data taking is then halted by disabling the clock trains during the 500 $\mu$s helicity-stabilization period. During this break 
between two MPS periods, the scaler data are latched into on-board memory in the scaler modules and the
scalers are then cleared.  Readout of the stored data via VME may then
proceed, even after the next MPS has begun, since readout of the stored
values does not affect the accumulation of data by the scalers. 

\subsection{Systematics Control Features}

The data-input latch is reset at the end of any clock train during which it was set. A more deterministic deadtime is achieved by 
using the NPN technique of disabling data from that channel during the next beam pulse. This is done by holding the reset, 
preventing the latch from being set, throughout the next clock train. Thus a calculable deadtime is associated with all recorded 
hits. Because the enforced deadtime is longer than that of the PMTs, discriminators or
meantimers, the resulting deadtime correction factor is independent of the less
well-defined recovery times of those devices.  Additional conventional deadtime correction
factors are required, however, for events that fire only some of the four
discriminators required to set the data-input latch.  Those
``singles'' corrections have a dependence on the deadtime of the discriminators. 

\begin{figure}[h]
\begin{center}
\includegraphics[width=13cm]{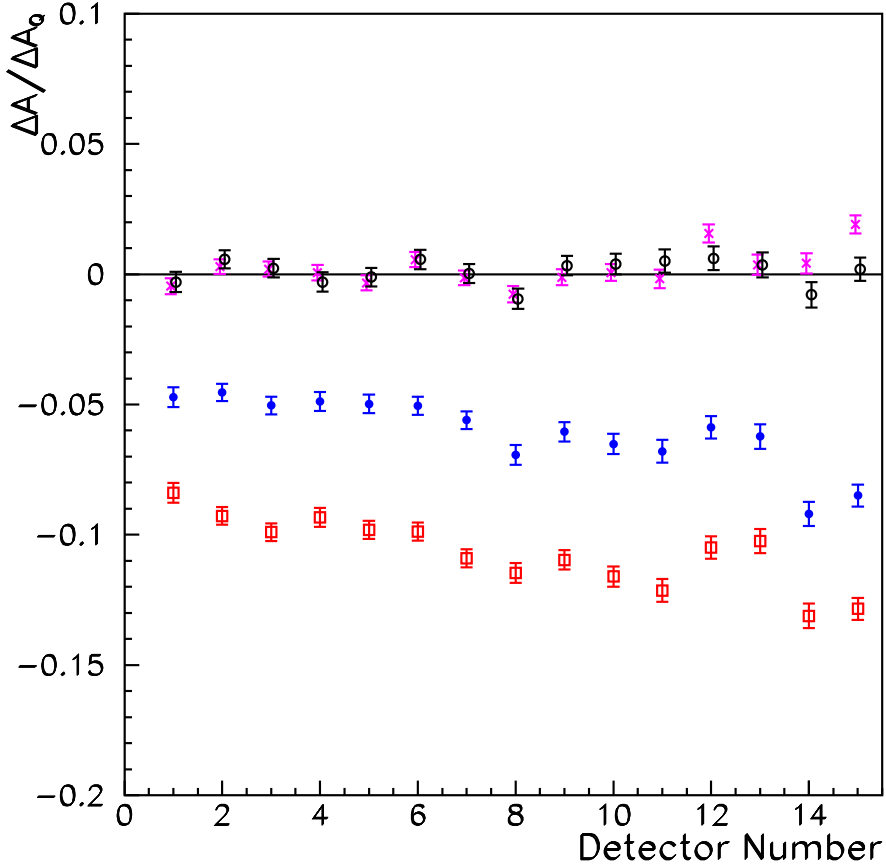}  
\vspace{0.5cm}
\caption{\label{fig:na_wells}
Measured slopes are shown for ratio of elastic-peak yield asymmetry (A) to charge 
asymmetry (A$_Q$) (for test data taken with deliberately large
A$_Q$).  Slopes are shown for raw yields (square), yields corrected
for deadtime imposed by time encoding board (solid circle), yields
corrected for this and for deadtime due to singles (X), and yields
with additional corrections from linear regression analysis (hollow circle).}
\end{center}
\end{figure}

The effects of these deadtime corrections are demonstrated in figure
\ref{fig:na_wells}.  This shows results from dedicated test running in
which a charge asymmetry, A$_Q$ was intentionally imposed to give
different beam intensity (differing by a few thousand ppm) for the different
helicities.  Yields are corrected to first order for beam-charge
variation by dividing the observed number of detected counts by the
integrated beam charge measured in the corresponding MPS.  A residual
yield-asymmetry, A, will be observed if the variation in deadtime loss
is not properly accounted for.  Since yield asymmetry is expected to
be proportional to the charge asymmetry, to first order, 
figure \ref{fig:na_wells} shows the observed slope, $\Delta \mbox{A}/ \Delta \mbox{A}_Q$.

In the absence of any deadtime correction, the raw elastic-peak yield 
asymmetry is
seen to be about 10\% of the applied charge asymmetry.   Correcting
for the deterministic deadtime, based only upon the measured number of
counts in bins of the time spectra, reduces this effect by about
half.  FastBus data were used to determine the rates of all possible
types of partial events.  These `singles rates' included events such 
as a single-CFD, a single meantimer, or any possible combinations.
Correction for deadtime due to these singles rates was based upon
estimated effective deadtimes, taking into account the 32 ns structure
of the beam.  These corrections are seen to properly account for the
remaining deadtime effect at the 1\% level except in the last 
detectors.  Subsequent, unpublished analysis showed that somewhat superior parameterizations of the
singles deadtime could be obtained by Monte Carlo simulation of
the electronics chain, but the results shown in figure \ref{fig:na_wells} were those used in
the published analysis of the data \cite{G0_FW_Results}.  Linear
regression analysis, applied to the main data set, determined the
residual dependencies of counting rates on beam parameters.
Application of those corrections to this test data set is seen to reduce 
the yield asymmetry to less than 1\% of the charge asymmetry for all
detectors.   Since the actual charge asymmetry during data taking was
typically less than 1 ppm, the residual yield asymmetries, after these
corrections, were negligible.
 
Each LTD board has two separate data-input latches and encodes time-of-flight
for two separate FPDs, known as a buddy pair.  The signal distribution  was chosen
so that each LTD board encodes the time-of-flight for two corresponding
diametrically opposite detector pairs.  That is, the signals from any detector
pair in octant one go to the same LTD as do the signals from the same detector
pair in octant five while signals from octants three and seven are similarly paired. Separate scalers then kept
track of how often a good hit occurred for each detector while its buddy was
busy, as indicated by the status of the buddy's data-input latch.  This gives
a rough estimate of the number of events that each LTD channel missed
because it was busy, itself, when a good hit would otherwise have been
counted. Separate counts were kept for the number of times a hit was recorded
when the buddy was busy because of a hit in the previous beam pulse and of the
number of times a hit was recorded when the buddy also recorded a hit in the same
beam pulse. Since coincidences are quadratic in the rates, they are
expected to have the same dependence on instantaneous beam intensity as do
deadtime losses. A helicity-correlated difference in buddy rates
would indicate a helicity-correlated structure to beam intensity.  If such a
helicity-correlated intensity structure went undetected it could cause a false
asymmetry by causing unexpected helicity-correlation in the actual deadtime
losses which would not be compensated by deadtime corrections based on mean
rates.  Analysis of the buddy scalers found that most runs showed no
indication of helicity-correlated intensity fluctuations, beyond the
expected statistical error.   A few runs, which showed large
correlations, were excluded from analysis.
 
\section{French Time-Encoding Electronics}
\label{section:FR}
The French electronics is fully integrated and is therefore very compact. All the required functionalities (CFDs, MTs, coincidence 
logic, time encoding and histogramming) are implemented on one single module processing 32 PMT signals. This module is 
referred to as DMCH-16X which stands for 
Discriminators, Meantimers, time enCoding, Histogramming, 16 MT channels within vXi standard. Thus, the 256 PMT signals 
from the four French octants are processed using eight modules fitting in one C size 
VXI crate. This crate is adapted for the 650W of power dissipated by the modules. \\
An additional module, called Interface Box (IB), provides common 
signals such as Y$_0$ and MPS to the eight DMCH boards through the VXI back-plane.

\subsection{Interface Box}
\label{subsection:IB}
The Interface Box (IB) is a custom VXI device designed at IPN Orsay which controls all the DMCH-16X modules by means of 13 
bus lines on the backplane avoiding cabling on module front panels. The IB is controlled through 
Device Dependent Registers. During the experiment the IB provides external signals (Y$_0$, MPS, 120 Hz) coming from 
the accelerator and the DAQ system to the DMCH-16X modules. Under the
control of the Trigger Supervisor the IB enables the data
collection. For test purposes of the French TEE subsystem only,
independently from the Trigger Supervisor, the IB generates its own
simulated signals (Y$_0$, MPS, 120 Hz) with accurate timing from a 31.2 MHz
internal crystal oscillator using an internal sequencer.
The Interface Box also makes available on its front panel an internal
generator signal synchronized to the Y$_0$ signal, to be used to test
the Constant Fraction Discriminators and the meantimers of the
DMCH-16X modules.

\subsection{DMCH-16X mother board}
\label{subsection:DMCH}

\begin{figure}[h]
\begin{center}
\includegraphics[width=13cm]{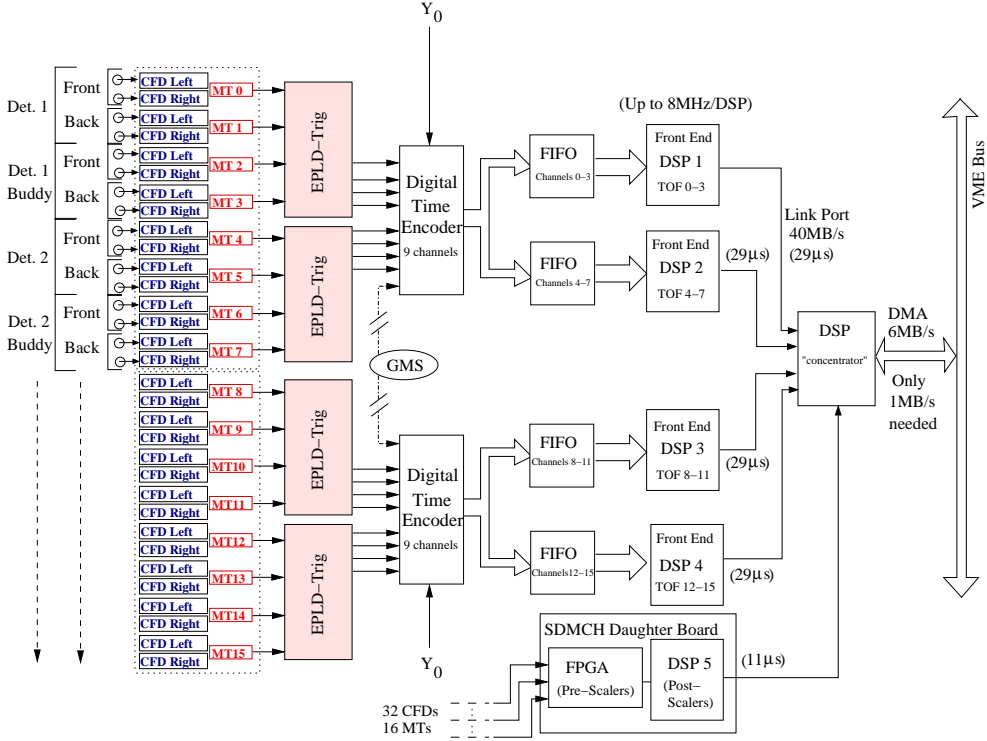}  
\vspace{0.5cm}
\caption{DMCH architecture. The DMCH-16X mother board processes 32 PMT signals 
corresponding to eight pairs of scintillators. The scaler daughter board (S-DMCH) 
holds one FPGA (Field Programmable Gate Array) chip (104 scalers) and
one Digital Signal Processor (DSP). 
To handle the 256 PMT signals of the four French octants, eight boards are required.}
\label{fig:dmch_architecture}
\end{center}
\end{figure}

Each DMCH module processes 32 PMT signals issued from eight detectors. The front-end electronics consists of 16 
daughter boards, each holding two Constant Fraction Discriminators (CFDs) and one meantimer (MT), 
both of analog design. The CFD-MT daughter board is discussed in section 
\ref{subsubsection:CFD_MT} as is the additional 
generator daughter board used to internally test the CFDs and MTs. As shown in figure 
\ref{fig:dmch_architecture}, the inputs of the CFDs are the right and the left PMT signals of each
scintillator. Each MT signal corresponds to a front or a back scintillator. To 
reduce time differences due to signal transit time inside the electronics circuit, the front and the back 
channels of any one detector are implemented next to each other. The coincidence between the front and back 
MT signals as well as other functions, such as the buddy features (`buddy' or `differential buddy') 
are implemented in an EPLD-Trig (Electrically Programmable Logic Device for Triggering) chip. Each chip handles one detector 
and its buddy detector (four MT channels).

\begin{figure}[h]
\begin{center}
\includegraphics[width=13cm]{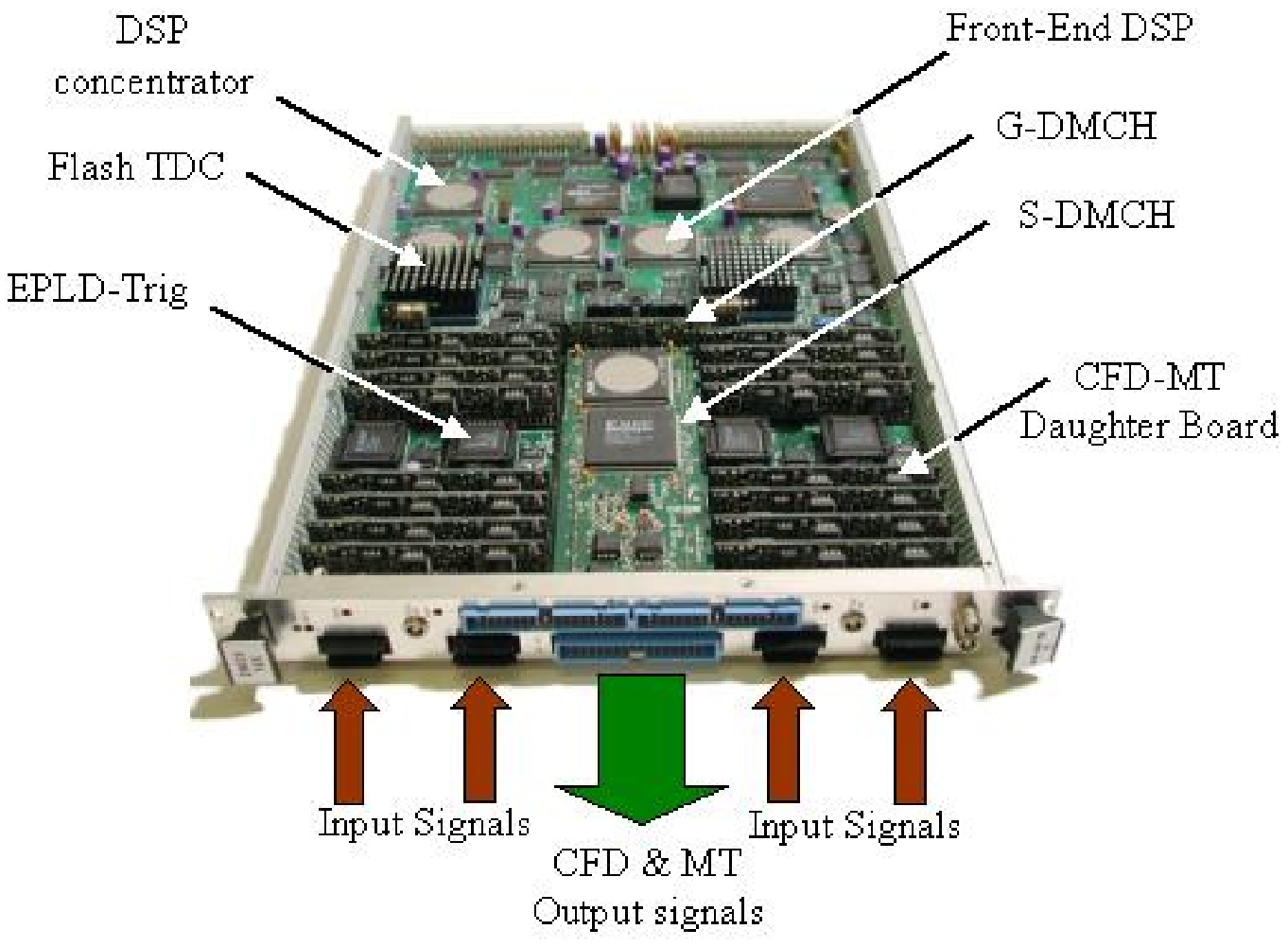}  
\caption{Opened view of the DMCH-16X mother Board with its three types of daughter boards (CFD-MT, G-DMCH 
and S-DMCH holding a FPGA chip and a DSP) and its main components: four EPLD-Trig, two 
digital time encoders (flash TDC), four front-end 
DSPs and one DSP concentrator.}
\label{fig:photo_dmch}
\end{center}
\end{figure}

In total there are four EPLD-Trig chips on each DMCH board to treat four detectors from two octants symmetrical 
to each other at 180 degrees w.r.t. the  beam axis. The coincidence window width was set to 7 ns. 
The opening of the coincidence window is driven by the arrival of an MT signal from the back scintillator.
For that purpose the front PMT signal was delayed by 17 ns by increasing the length of the signal cables.
In case of a front-back coincidence event the arrival time of the MT front signal is tagged by the fast digital Time Encoder (TE) 
relative to the Y$_0$ beam pick-off signal. The principle of this TE will be described in section \ref{subsubsection:CTN}. One of 
the main advantages of this TE, when compared to the 1 ns NA electronics, is its time resolution of 250 ps which makes the 
background subtraction easier. As each TE holds eight channels available to flash the time, two TEs sit on a DMCH mother 
board implying a symmetry that can be seen in figures \ref{fig:dmch_architecture} and \ref{fig:photo_dmch}. 

\begin{figure}[h]
\begin{center}
\includegraphics[width=10cm]{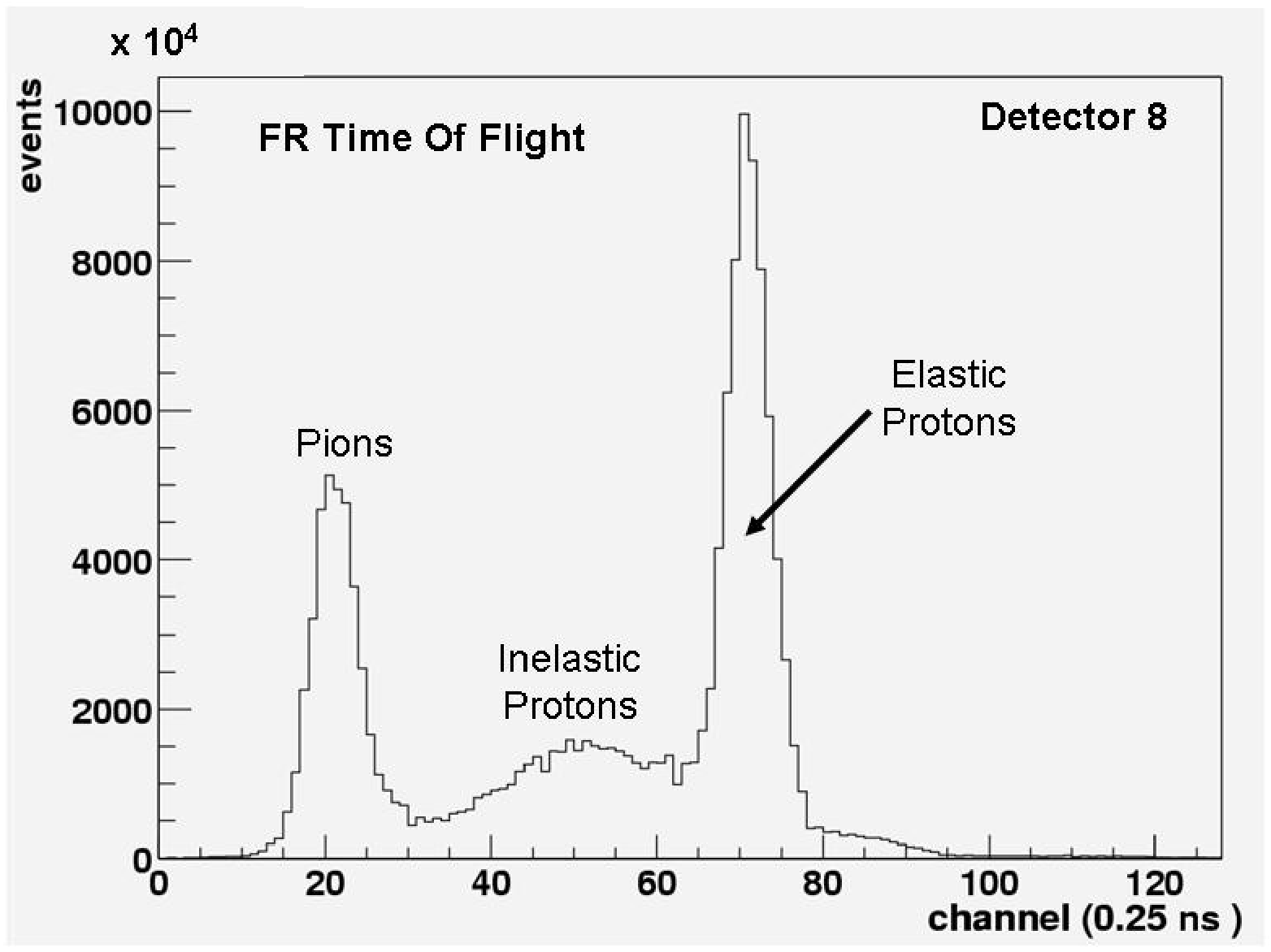}  
\caption{Resulting time-of-flight spectrum for Detector eight (middle of the focal plane) built by the FR electronics (128 time bins 
of 250 ps). This ToF spectrum corresponds to an accumulation of 18274 MPS (10 minutes of data taking)}.
\label{fig:TOF_Det8}
\end{center}
\end{figure}

The event-time information from 
each TE is transferred to one of the two DSPs 
associated with the TE using asynchronous FIFOs (2048 words), see the DMCH architecture presented in figure 
\ref{fig:dmch_architecture}. Each DSP builds two Time-of-Flight 
(ToF) histograms from coincidence hits collected during one MPS corresponding to 
two detectors (one detector and its buddy). 
Each such spectrum has a fixed length of 256 Bytes.
An example ToF spectrum is presented in figure \ref{fig:TOF_Det8}.

Histograms stored in the four Front-End (FE) DSPs (ADSP-21062 SHARC) are then 
transferred to a DSP concentrator (ADSP-21062 SHARC) after the end of each MPS. One of the main challenges was to 
be able to build ToF spectra at such a high rate and to transfer them from the four FE DSPs to the DSP 
concentrator after the end of an MPS during the helicity-stabilization
period. This has been achieved by using the chosen 
type of DSP which handles 40 MIPS and which has 2 Mbit of on-chip memory. The code of the FE DSPs has 
been optimized so that in only four instructions (five cycles of 25 ns) one bin is incremented. 
Therefore the maximum rate one FE DSP can handle is 8 MHz (4 MHz per detector). Before 
the transmission of ToF spectra to the DSP concentrator, 66 kbit of memory is used per FE DSP to 
store all the different spectra. The fast transfer from the FE DSPs to the DSP concentrator relies on link ports 
(40MB/s) which work in parallel. This transfer takes 29 $\mu$s.
  
Besides the time encoding system, each DMCH board holds a daughter board, referred to as an S-DMCH (S- standing for 
Scalers), consisting of one FPGA chip and a DSP (see section \ref{subsubsection:SDMCH}). 
This DSP collects, for each MPS, the number of counts in each CFD 
and MT. In total 104 scalers are accessible. As 
for the four front-end DSPs, the data from the S-DMCH DSP are transferred to the DSP concentrator at 
the end of each MPS. The data transfer (1264 words of four Bytes per DMCH module per MPS) to the  
Read Out Controller (ROC) occurs during the next MPS by 
Direct Memory Access (DMA). At a rate of 6 MB/s, it takes 16 ms. From the ROC all the DMCH data 
are then gathered with the data from the other ROCs in the event builder. 
The total rate of the French data is 1.4 MByte/s (46 kByte/MPS) which represents 2/3 of the total G$^0$ data transfer.

From the CFD-MT to the fast TE, all the circuits use ECL technology whereas 
the DSPs, the FIFOs and the VXI management are based on MOS technology.

\subsubsection{CFD-MT Daughter board:}
\label{subsubsection:CFD_MT}

\begin{figure}[htbp]
\begin{center}
\includegraphics[width=13cm]{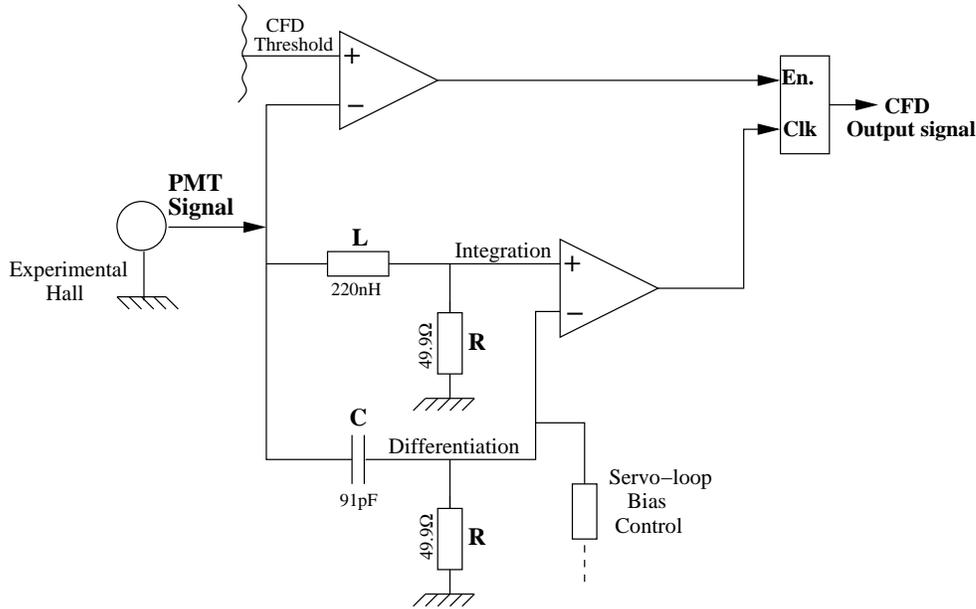}  
\caption{CFD schematics.}
\label{fig:CFD_schematics}
\end{center}
\end{figure}

The CFD-MT daughter board holds two analog constant fraction discriminators, a coincidence circuit and an analog 
meantimer whose output delay is software-adjustable.
The two discriminators accept the analog pulses from the photomultiplier bases. 
The advantage of constant fraction discriminators is that the firing time is independent 
of the amplitude of the input signals. 
Due to the necessity of space-saving to fit within a VXI module, a pulse shaping, integration-differentiation 
type CFD, using small volume L, R, C components has been chosen instead of the usual bulky delay-lines. This shaping 
allows the CFD input impedance to be purely resistive, not frequency dependent, provided the condition RC=L/R 
is achieved. A good quality 50 $\Omega$ termination is obtained for the long PMT signal cables.
The schematic of this type of CFD is presented in figure \ref{fig:CFD_schematics}.

The cross-talk between two adjacent channels has been measured. For an input signal of 3 ns rise time and 1 V pulse height, 
the cross-talk is -55 dB corresponding to an induced signal of 1.5 mV in the adjacent channel. 
The CFD thresholds are software-selectable from 0 mV to 255 mV (in 1 mV steps). 
During G$^{0}$ runs the CFD thresholds were set to 50 mV.\ \ It has been shown, using a simulation, that no pile-up effect could 
be observed under the conditions of G$^0$.

\begin{figure}[h]
\begin{center}
\includegraphics[width=11cm]{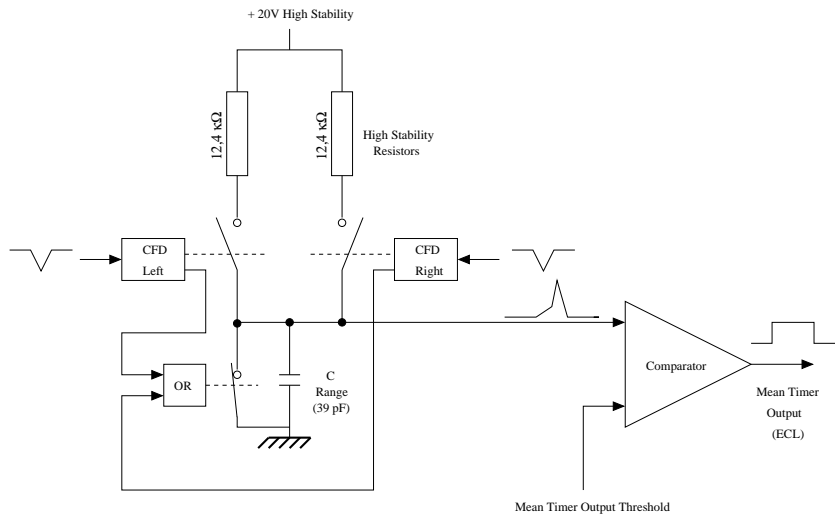}  
\caption{Schematics of the meantimer (waiting state).}
\label{fig:schema_MT}
\end{center}
\end{figure}

\begin{figure}[htbp]
\begin{center}
\includegraphics[width=11cm]{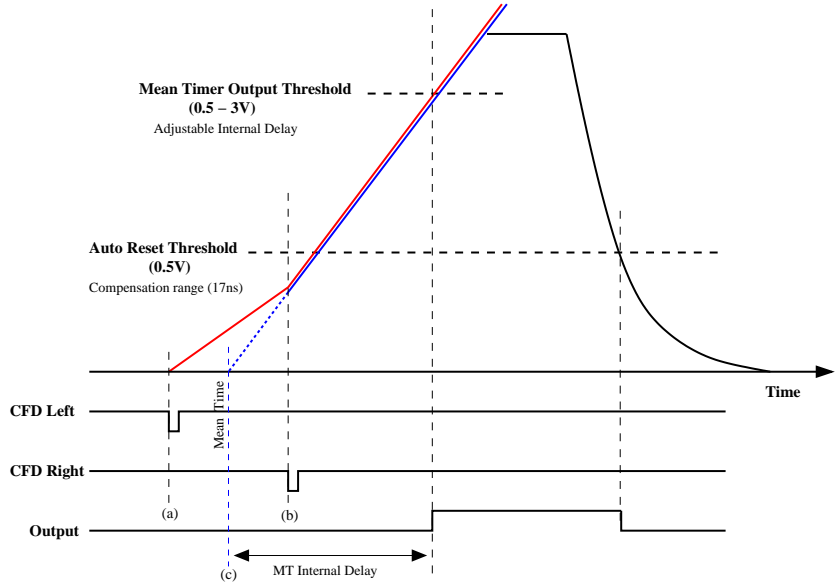}  
\caption{MT Principle.}
\label{fig:MT_principle}
\end{center}
\end{figure}

After an input signal, the discriminator is disabled until the end of the meantiming sequence. If only one
CFD pulse (left or right) is present during the MT compensation range, the sequence is reset. 
The MT compensation range, chosen to match the maximum transit time associated with the longest scintillator 
paddle has been set to 17 ns, by the value of the capacitor used. The schematics of the meantimer and its principle of operation 
are described respectively in figures \ref{fig:schema_MT} and \ref{fig:MT_principle}. 

The MT is based on a principle which has been used at CERN since the 1960's but 
the design and  implementation originates from the electronics department of IPN Orsay. 
Once either discriminator is fired, a capacitor is 
linearly charged with a constant  current, when the second CFD fires a second equal current is 
added so as to make the voltage slope twice as fast. 
An output signal is generated as soon as the output threshold is reached. This output threshold,
corresponding to the MT output delay, is adjusted by software and is mainly used to tune
 the coincidence between the front and the back signals. The MT output delay can be set from 
0 to 44 ns in 0.175 ps steps.     
To avoid any time distortion on the front MT signal, only the back MT output 
delay has been adjusted while the front PMT signals have been delayed by 17 ns in hardware.
If only one discriminator gets a signal, then the meantimer is reset after the charge reaches the MT 
auto reset threshold.
The MT deadtime associated with a complete MT event (two CFD pulses within the compensation range) is of the 
order of 37 ns depending on the setting of the back MT output delay.
If only one discriminator gets a pulse then the associated MT dead time is increased to 40 ns because 
the MT reset occurs after the fixed compensation range.  
The stability versus temperature of the meantime depends only on the stability 
of two resistors and on the capacitor used for the compensation range. Since low temperature-coefficient 
components are used, the resulting drift is expected to be in the range of a few ps/$^{\circ}$C.
The cross-influence between two adjacent MT channels results in a 60 ps shift.
 
\subsubsection{Internal Generator Daughter Board}

The G(enerator)-DMCH is a custom generator designed to test the function of the 
discriminators and meantimers circuits. It is implemented on an additional daughter board 
plugged onto each DMCH-16X mother board. The G-DMCH provides two output signals simulating the left and the right
PMT signals. The separation in time between the 
two pulses varies from 0 to the maximum time difference (17ns) corresponding to the total length of the 
longest scintillator. Furthermore, the amplitudes of the two pulses are software adjustable to check the 
discriminator threshold values.

\subsubsection{Scaler Daughter Board (S-DMCH):}
\label{subsubsection:SDMCH}

In the original design of the DMCH-16X module, this daughter board
held only the 32 CFD outputs to make 
them available on the front panel of the DMCH module to be sent to the FastBus TDCs.
Soon, it became apparent that it was very important to be able to
quantify individually the number of CFD and MT signals in order 
to ease the evaluation of the dead time, particularly that associated with incomplete events 
(less than four CFD signals per detector). As the surface of the S-DMCH daughter board was fixed, 
it was decided to use a system of pre- and post-scalers, as described below.

For each MPS, the total number of events in each of the 32 CFDs and the 16 MTs is recorded independently of the arrival time of 
a hit and the completeness of the event (coincidence MT front-back). Moreover to get time information of the CFD and MT 
signals, a sliding time gate is used  to count the number of CFDs and MTs 
occurring within a 2 ns time bin for each MPS. This 2 ns time window slides by 2 ns at each MPS to scan 
the entire 32 ns period used to build ToF spectra. The last eight scalers are 
the `buddies' explained in section \ref{subsection:TEE}.
The CFD and MT scalers from S-DMCH are complementary to the FastBus system and have been very useful in evaluating
the dead time arising from events corresponding to only one CFD or one MT.

The 104 scalers are implemented using two chips: a FPGA (XILINX) supplying 104 pre-scalers (seven bit) 
and a DSP (ADSP-21062 SHARC). The DSP supplies 104 post-scalers, each of them being incremented by 127
at each overflow of the corresponding pre-scaler. In fact, once a pre-scaler 
reaches 127, it overflows and its corresponding carry bit is set to one. 
The pre-scaler itself is reset and keeps on counting. The DSP loops over 
the 104 pre-scaler in 23.5 $\mu$s to look for carry bits which are set to one. At the end of each MPS, the DSP
 reads the carry bit and the remainder of each pre-scaler and transfers the content of the 104 post-scalers 
to the DSP concentrator implemented on the DMCH mother board.
As each pre-scaler has only seven bits for counting and a carry bit, it might happen 
that a pre-scaler, especially the ones corresponding to CFDs, overflows over 255 (carry bit already set to 1)
 before being read by the DSP. To keep track of such possible overflows, each pre-scaler has an extra 
bit referred to as the overflow bit. When the S-DMCH DSP reads a pre-scaler, if the carry bit and the overflow bit are set to one, 
then the DSP increments one specific memory location. At the end of each MPS, only the total number of overflows per S-DMCH 
is known. As it is not possible to correlate the overflows with individual pre-scalers, no correction 
can be made on the event counting over 255 but this global information has been  used as an overall alert 
which was very useful to detect any variation in the event rates because in normal beam conditions 
no overflows occur. Therefore, in the off-line analysis, any quartets for which one or more alerts were found were 
discarded.

\subsubsection{Digital Multichannel Time Encoder}
\label{subsubsection:CTN}

For the G$^0$ experiment, which is based on time-of-flight measurements, the time encoder 
is the central component as it provides the arrival time of an event. The best
achievable time resolution is advantageous to achieve better background subtraction. 
In the French design, the digital multichannel time encoder 
provides time bins of 250 ps. This was designed by the electronics department of 
the Institute of Nuclear Physics in Orsay (France). Independently from G$^0$ requirements, the 
motivation for the design and the development of this ASIC, based on 1.2$\mu$ ECL bipolar prediffused 
technology, started in 1990 to provide a generic tool for nuclear 
physics experiments relying on accurate time measurements. This TE has
nine independent channels. The ninth
 channel, which serves as a start signal wasn't used for G$^0$. Instead, the start for the eight 
channels is given by a Voltage Controlled Oscillator (VCO) locked on the Y$_0$ reference signal. 
Consequently, the encoded time corresponds 
directly to the ToF without any calculation. The possibility of using this mode for G$^0$ is 
one of the main features of the time encoder which, coupled to FIFOs and front-end DSPs, 
allow us to build ToF spectra at a rate of 4 MHz per detector.

\begin{figure}[htbp]
\begin{center}
\includegraphics[width=10cm]{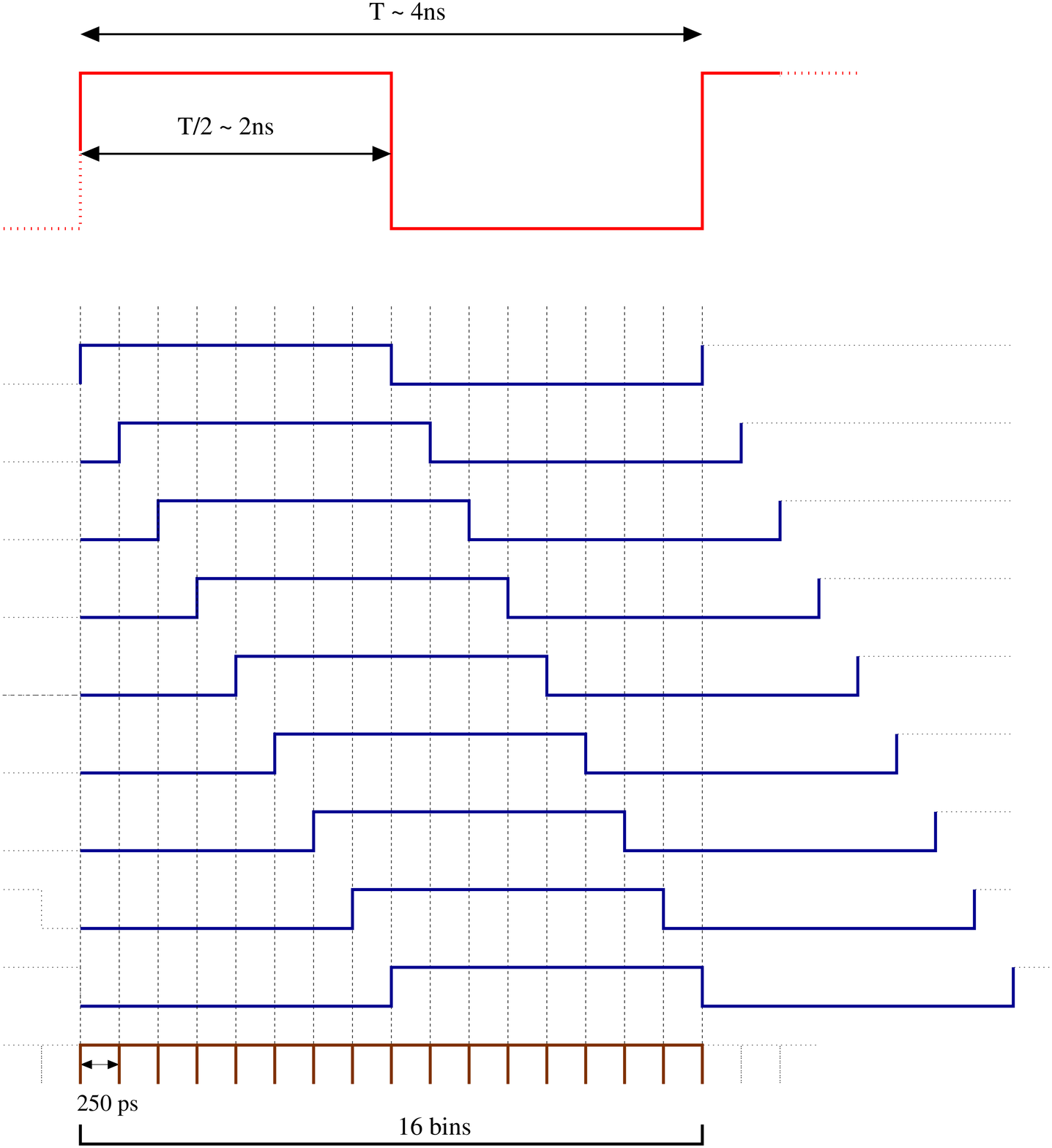}  
\caption{Basic principle of the period slicing using the reference signal ($\approx$ 4ns) using eight Delay Locked Loop (DLL) 
circuits.}
\label{fig:DLL}
\end{center}
\end{figure}

\begin{figure}[htbp]
\begin{center}
\includegraphics[width=15cm]{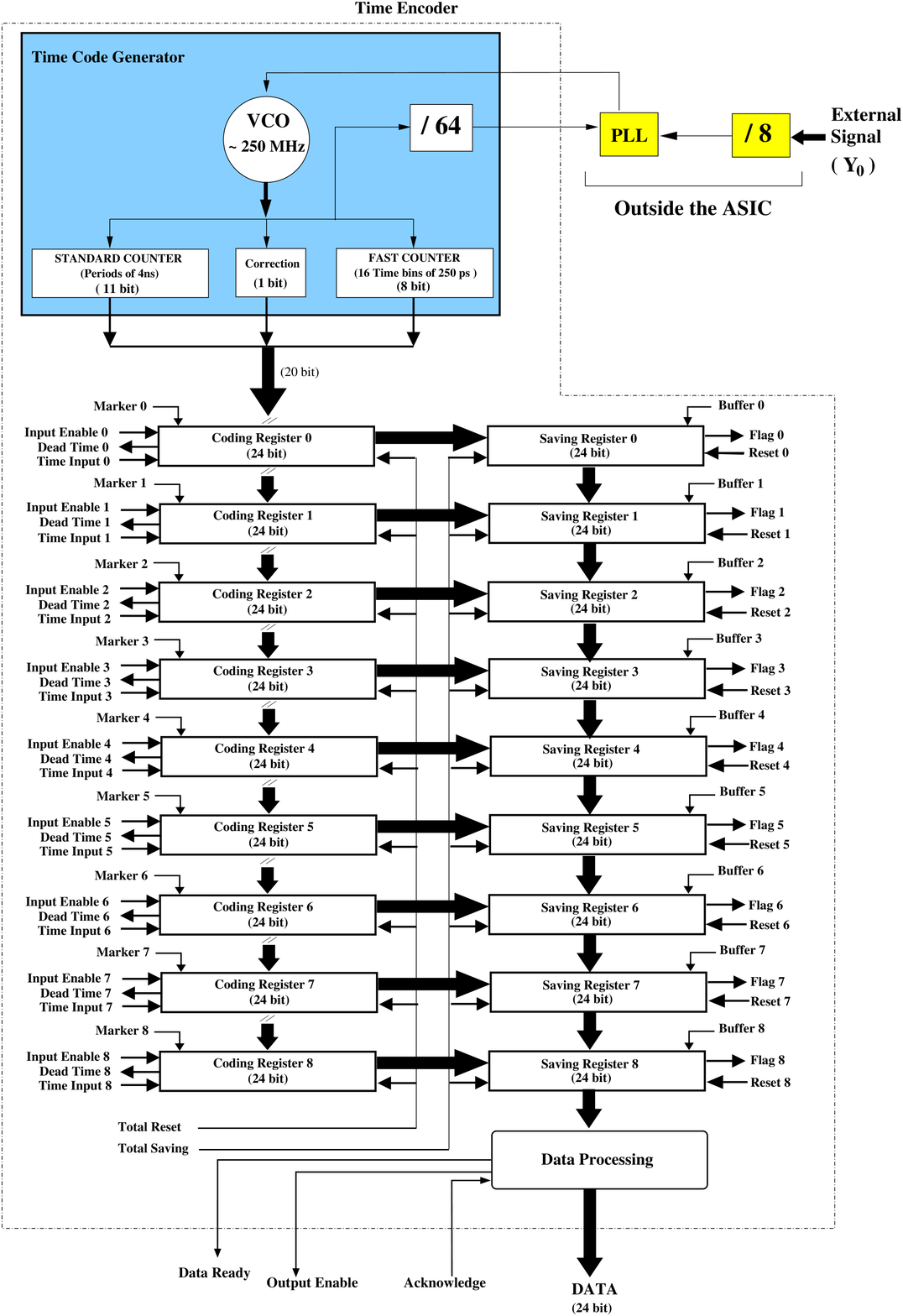}  
\caption{Internal architecture of the Time Encoder. The nine inputs to coding registers are independent channels.
As indicated, the external Phase-Locked Loop (PLL) adjusts the internal Voltage Controlled Oscillator (VCO) to synchonize it to 
the Y$_0$ signal.}
\label{fig:CRT}
\end{center}
\end{figure}

The principle of the time encoding is very classical. Fundamentally, the TDC is made of a
crystal oscillator, delivering a period $T$ (here $\approx$ 4 ns), 
a `standard' counter running at the frequency $\frac{1}{T}$ for counting of the periods and a `fast' counter, slicing
the period $T$ (interpolation). The standard counter runs continuously. The fast counter relies on Delay Locked Loop (DLL) 
circuits which split the period of the standard counter into sixteen time slices of $\approx$ 250 ps
representing the time bins, see figure \ref{fig:DLL}. The frequency of the standard counter is provided by a VCO (figure 
\ref{fig:CRT}). 
This frequency {\it{f}}$_{VCO}$ divided by 64 is locked on an external frequency {\it{f}}$_{ext}$. 
In the case of the G$^0$ experiment, the external frequency is the Y$_0$ signal of 31.1875 MHz divided-down 
by eight to match with {\it{f}}$_{VCO}$ / 64. This comparison is made by a Phase Locked Loop (PLL) 
which generates a proportional error voltage to continuously adjust the VCO frequency to the external 
frequency.

The total width of the first eight time slices is locked to half of the VCO period, 
see figure \ref{fig:DLL}. As the width of each individual time slice is not locked and 
because the transit time from one transistor to another can differ, a time differential 
non-linearity from bin to bin can be observed. 

According to the scheme presented in figure \ref{fig:CRT}, the encoded time results from a fine and a coarse time code.
The fine time code (0 through 15) given by the period-slicing corresponds to the time 
within one period ($\approx$ 4 ns). This fast counter is a 
Johnson-type counter requiring only one bit to change at a time. 
The coarse time code provided by the standard counter tells us in which 
period the event occurred among the eight periods required to get the Y$_0$ period. 
This standard counter is a natural-progression binary type. 

In the design of such a Time Encoder,  three issues require special care.
First, as the arrival of the physics event is asynchronous with the internal frequency of the Time Encoder, care must be taken to 
ensure that the time code is unambiguous when  the latch of the fast counter occurs. This has been solved by using a Johnson 
type counter. Second, the standard counter being a conventional binary counter driven by the internal frequency of the Time 
Encoder, has to be latched when its contents are stable. To achieve this requirement, the triggering of the latch of the standard 
counter is delayed with respect to the internal frequency of the Time Encoder such that it occurs in a time window in which all the 
bits are stable. Lastly, the standard counter code has to be correctly aligned with the code provided by the fast
counter. In practice, it is impossible to increment the standard counter
exactly in coincidence with the roll-over of the fast one. Because the triggering of the latch of the standard
counter cannot be perfectly synchronized with the latch of the fast counter, an error of one period $T$ could occur. This
error is corrected by means of a supplementary `corrector' bit which
stores the value of the least significant 
bit of the standard counter at the middle of the period of the fast
counter. When the standard counter is latched, the value 
of its least significant bit is compared to the corrector bit and a table is used to determine whether a correction of the least 
significant bit is to be made within the ASIC itself. 

Once an event occurs in any of the eight channels of the TE, the contents of the two counters are latched
into the corresponding coding register. The nine coding registers are independent of each other. Each 
coding register is linked with a saving register in order to quickly release the coding register, lowering the coding dead time (24 ns). The coding registers are arranged in a synchronous FIFO architecture 
to buffer a possible burst of input signals as well as to multiplex and send the coding 
results. 

The intrinsic dead time of the Time Encoder is of the order of 24 ns.
Compared to the 37 ns of dead time associated with the meantimers, the TE does not add any extra dead time.

Intrinsically, the time encoder presents a Differential Non Linearity (DNL) from bin to bin which 
is very sensitive to temperature variations. For a given temperature of operation, the DNL of the time encoder can be minimized 
by software. As the DNL acts on the bin width, an off-line correction has to be applied 
to allow correct extraction of  the events associated with the different Q$^2$ from the same detector 
(as occurs for detectors 14 and 15 of each octant, as explained in
section \ref{section:Q_bins}). Periodically, in the absence of beam,
background data were taken using a radioactive source. 
Ideally, for long data-taking with no DNL effect, we would expect flat time-of-flight spectra. In fact, 
without beam, the deviations seen in the ToF spectra compared to a flat spectrum are a measure of the 
differential non linearity of the time encoder. Relying on the background ToF spectra, the width of 
each bin is adjusted by software and a corrector factor is applied to
the ToF obtained with beam. More details 
on the principle of this correction can be found in reference \cite{these_Guillaume}.

Independent of the DNL, the ToF spectra present a bin-to-bin correlation. This effect has been well reproduced by a simulation 
(Appendix A of reference \cite{these_Silviu}) which indicated a jitter of 50ps modulated at 1 to 10 kHz. This jitter is strongly 
suggested to originate from electronic noise (voltage control of the TE, etc.) surrounding the DLL circuits which build the time 
bins at the level of the fast counter. This effect is amplified by the implicit mode of operation chosen for G$^0$. The global 
systematic uncertainty due to the bin-to-bin correlation has been estimated to 0.025 ppm.

\subsection{Dead Time corrections}
One significant issue for the $G^0$ experiment is the ability to deal with the deadtime of the electronics. If associated with beam 
charge asymmetries and/or detector counting asymmetries for time ranges outside of the one of interest, the deadtime could 
induce sizeable asymmetries on the elastic counting rates. The resulting systematic
error on the asymmetries, which is proportional to the deadtime itself, will then be minimized by
correcting for the deadtime losses as well as possible. It is then important to know the precision
that can be achieved for deadtime corrections. 

As discussed above the deadtime is due to the front-end electronics,
upstream of the time encoding system, consisting of the discriminators (CFD) and meantimers (MT). Also to improve 
the accuracy of the correction, the Next Pulse Neutralization (NPN) procedure has been implemented,
meaning that the deadtime is extended precisely up to the end of the
next 32 ns time window, for coincidences between the front and back meantimers. 

Due to the beam time structure, the deadtime correction, to be applied to the measured counting rates 
before calculating the asymmetries, must be estimated bin by bin over the 32 ns time spectra. 
This consists of two main contributions. 
The first is due to the probability for having a preceding coincidence between the back and the front detectors, 
which prevents encoding of an event at bin $i$ (32 ns corresponds to 128 bins of 250 ps width). 
This probability can be determined directly from the time-of-flight spectra provided by the DMCH16X,
including the effect of the NPN system.
The second contribution to the deadtime, is due to the probability for one single meantimer (front or back) 
or one of the four discriminators being fired, which can prevent a
subsequent front-back coincidence from being encoded.
The time distributions associated with single CFDs or MTs are obtained from the analysis of the FastBus data,
which also allows proper accounting for all possible configurations, such
as firing of two CFDs which are  
not associated with the same meantimer or firing of one meantimer in coincidence with one single CFD. 
As discussed in section \ref{subsubsection:CFD_MT}, the deadtime due to the meantimer itself, 
differs when only one or both CFDs fire(s) on the same meantimer.
The deadtime correction was calculated by taking into acount the
appropriate deadtime for each of the different configurations
with the normalization being provided by the individual scalers of all CFDs and meantimers implemented on 
the daughter board S-DMCH.

A number of studies have been carried out to check the ability to
accurately correct for deadtime losses.

A first test, dedicated to the deadtime correction due to  front-back coincidences, used
a high intensity Strontium source coupled to a PMT, allowing counting rates up to 4 MHz.
The time distribution of the measured coincidences, relative to an
arbitrary 32 ns time window, depends upon
the time bin due to the NPN system, with the slope of the time distribution being directly proportional 
to the deadtime itself.
This test demonstrated \cite{G0_Arvieux} that deadtime corrections, involving only coincidence events, 
can be predicted with a precision of about 1\%.

\begin{figure}[h]
\begin{center}
\rotatebox{0.}{
\includegraphics[width=11cm]{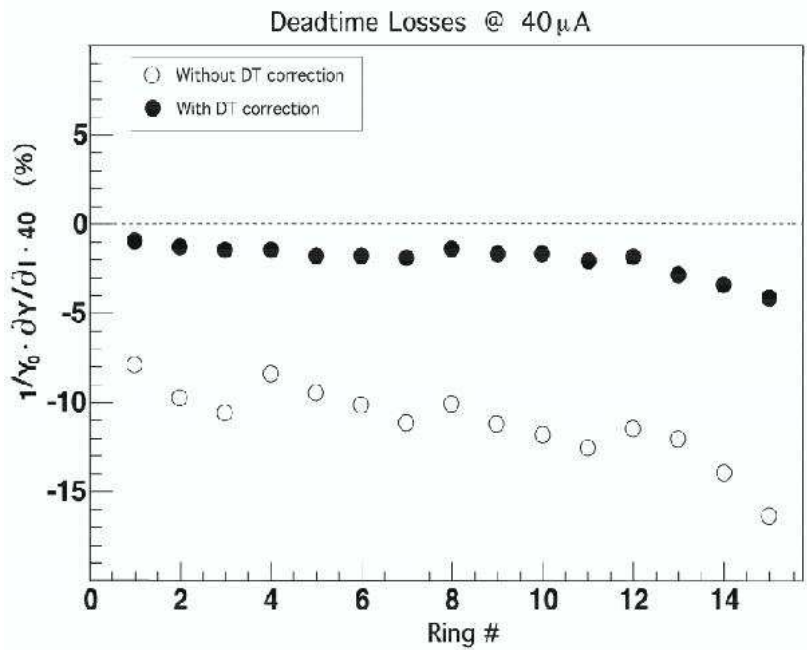}  }
\caption{Residual deadtime obtained from the dependence of the normalized yield on beam current. Here the observed effect of 
deadtime is plotted before and after deadtime correction for the French octants.}
\label{fig:SlopesNormLumi}
\end{center}
\end{figure}

A second test was carried out in the $G^0$ beam
conditions, where the deadtime losses
due to single CFDs and MTs were comparable to those from front-back coincidences \cite{these_Guillaume}.
Measurements were performed with several beam currents
up to 40 $\mu A$ \cite{these_Benoit}. The slope of the yield (normalized to the beam charge), 
against the beam current is expected to be proportional to the deadtime. This slope can be extracted 
before and after applying deadtime corrections and is used to estimate the precision of the correction.
Figure \ref{fig:SlopesNormLumi} shows this slope normalized by the
yield, itself, (in \% /$\mu A$) for each detector.  This has been
multiplied by the beam current to obtain the corresponding deadtime. It can be seen that 
the deadtime at the nominal beam current of 40 $\mu A$ ranges between
8 and 16\%.  Application of the 
deadtime corrections due to single CFDs, single MT and coincidences reduces the beam current dependency
to a residual deadtime of 1 to 4\% for the French detectors. 

\begin{figure}[h]
\begin{center}
\rotatebox{0.}{
\includegraphics[width=11cm]{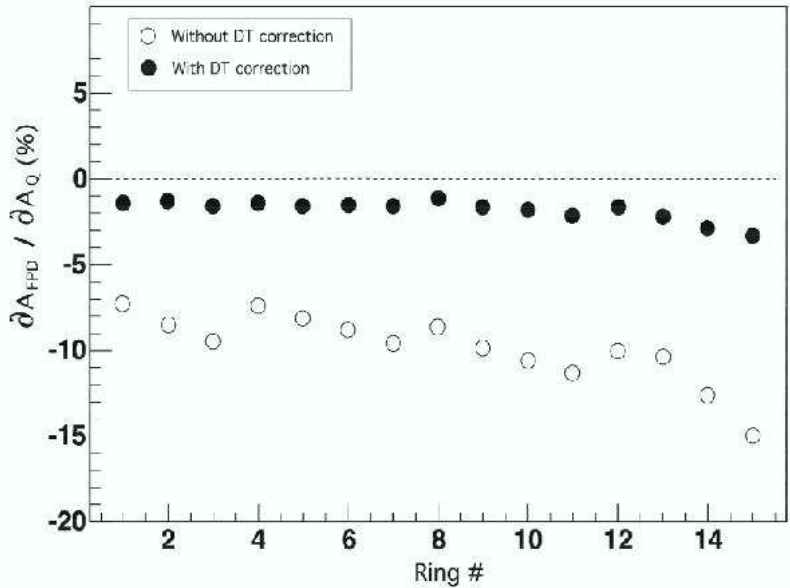}  }
\caption{Residual deadtime obtained from the dependence of the
  detector yield asymmetry as a function of beam charge asymmetry for different deadtime corrections. Here the deadtime is 
plotted before and after deadtime correction for the French octants. Linear regression analysis is not included.}
\label{fig:WellsPlot}
\end{center}
\end{figure}

Another study was performed by analyzing the detector yield asymmetry as a function of the 
beam charge asymmetry \cite{these_Benoit}. For this test, dedicated runs were recorded with charge asymmetry 
varying between -4000 and +2000 ppm, imposed by a Pockels cell device at the electron source. The slope
of the detector yield asymmetry versus the beam charge asymmetry
is directly proportional to deadtime. Figure \ref{fig:WellsPlot} shows the results obtained at 40 $\mu A$,
which are in agreement the previous ones obtained by varying the beam current between 0 and 40 $\mu A$.

In conclusion, the studies performed on the French electronics have
shown that we were able, in realistic conditions,
to estimate about 80 to 90\% of the deadtime using an analytical approach without any free parameters. 
The remaining dependence of the counting rates on the beam current was removed using a linear regression analysis.

\section{Summary}

\begin{figure}[h]
\begin{center}
\includegraphics[width=11cm]{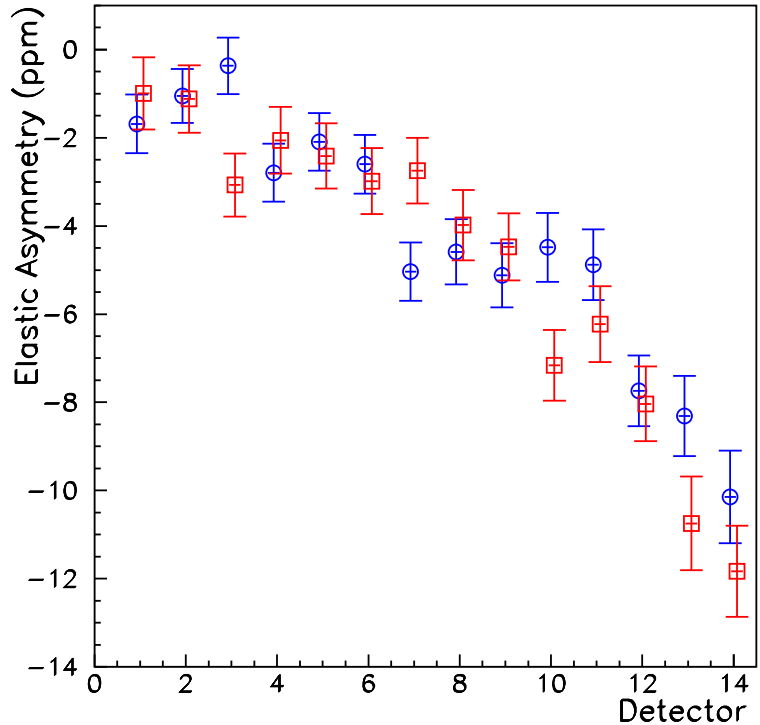}  
\caption{\label{na_fr}
Elastic asymmetries, detector-by-detector, as measured using only the
four octants instrumented with North American apparatus (circles) or
the four octants instrumented with French apparatus (squares).  Error
bars represent statistical errors.  Data points have been slightly
displaced horizontally to aid in comparison.}
\end{center}
\end{figure}

The forward-angle part of the G$^0$ experiment required the
equipment described here to permit the extraction of the
elastic-scattering asymmetry.  This part of the experiment has been
successfully carried out and the results of the analysis have been
presented in reference \cite{G0_FW_Results}.  That analysis
combined the results from all octants of the spectrometer.  Figure
\ref{na_fr} shows the elastic asymmetries, as extracted
detector-by-detector, separately for the four octants instrumented
with French electronics and detectors and for the four octants
instrumented with North American electronics and detectors.  
These asymmetries are extracted, for each detector, by correcting for
background contamination, as described in reference \cite{G0_FW_Results}.

No evidence is seen of a systematic difference when the results obtained
with one set of equipment are compared to the other.  Despite
significant differences in design and operation, the results from the
two sets of apparatus are seen to be in excellent agreement.  This
serves to strengthen confidence that neither set of apparatus has
introduced unintended systematic distortion of the data.

\section{Acknowledgments}
This work was funded in part by the U.S. Department of Energy under Contract DE-FG02-87ER40315 and by CNRS/IN2P3.\\

We gratefully acknowledge the strong technical contribution to the design, the implementation and the resolution 
of problems related to the French electronics from the electronics department of IPN Orsay namely: 
A. Amarni, D. Allibert-Rougier, J.-C. Artiges, A. Bouricha, V. Chambert, P. Cohen, D. Desveaux, 
P. Edelbruck, D. Lalande, S. Pr\'e, S. Tanguy, also from the Orsay computer group namely J.-F. 
Clavelin, P. Gara and H. Harroch. Many thanks to D. Abbot and S. Wood from JLab for their help in resolving DAQ 
problems. We would like also to acknowledge the technical staff from LPSC which have contributed to the DAQ and electronics
implementation : G. Barbier, G. Bosson, D. Dzahini, J.P. Girard, S. Muggeo, J.P. Scordilis.

\end{document}